\input harvmac.tex
 \input epsf.tex
 \input amssym

\def\figin{\epsfcheck\figin}\def\figins{\epsfcheck\figins}
\def\epsfcheck{\ifx\epsfbox\UnDeFiNeD
\message{(NO epsf.tex, FIGURES WILL BE IGNORED)}
\gdef\figin##1{\vskip2in}\gdef\figins##1{\hskip.5in}% blank space instead
\else\message{(FIGURES WILL BE INCLUDED)}%
\gdef\figin##1{##1}\gdef\figins##1{##1}\fi}
\def\DefWarn#1{}
\def\figinsert{\goodbreak\midinsert}
\def\ifig#1#2#3{\DefWarn#1\xdef#1{fig.~\the\figno}
\writedef{#1\leftbracket fig.\noexpand~\the\figno} %
\figinsert\figin{\centerline{#3}}\medskip\centerline{\vbox{\baselineskip12pt
\advance\hsize by -1truein\noindent\footnotefont{\bf
Fig.~\the\figno:} #2}}
\bigskip\endinsert\global\advance\figno by1}

\def\unit{\relax{\rm 1\kern-.26em I}}
\def\nada{\relax{\rm 0\kern-.30em l}}

%\draftmode

\def \la {\langle}
\def \ra {\rangle}
\def \pa {\partial}

\def \eps {\epsilon}

%\def\Omega{\rho,\sigma,\nu  }

%% MACROS
\noblackbox
\def\IL{\relax{\rm I\kern-.18em L}}
\def\IH{\relax{\rm I\kern-.18em H}}
\def\IR{\relax{\rm I\kern-.18em R}}
\def\IC{\relax\hbox{$\inbar\kern-.3em{\rm C}$}}
\def\IZ{\relax\ifmmode\mathchoice
{\hbox{\cmss Z\kern-.4em Z}}{\hbox{\cmss Z\kern-.4em Z}} {\lower.9pt\hbox{\cmsss Z\kern-.4em Z}}
{\lower1.2pt\hbox{\cmsss Z\kern-.4em Z}}\else{\cmss Z\kern-.4em Z}\fi}

%% MORE MACROS

\font\manual=manfnt \def\dbend{\lower3.5pt\hbox{\manual\char127}}

\def\IZ{\relax\ifmmode\mathchoice
{\hbox{\cmss Z\kern-.4em Z}}{\hbox{\cmss Z\kern-.4em Z}} {\lower.9pt\hbox{\cmsss Z\kern-.4em Z}}
{\lower1.2pt\hbox{\cmsss Z\kern-.4em Z}}\else{\cmss Z\kern-.4em Z}\fi}

\def\bar{\overline}

\def\pa{\partial}

\def\rt2{\sqrt{2}}
\def\irt2{{1\over\sqrt{2}}}

\def\hat{\widehat}
%  \slashchar puts a slash through a character to represent contraction
%  with Dirac matrices. Use \not instead for negation of relations, and use
%  \hbar for hbar.
\def\slashchar#1{\setbox0=\hbox{$#1$}           % set a box for #1
   \dimen0=\wd0                                 % and get its size
   \setbox1=\hbox{/} \dimen1=\wd1               % get size of /
   \ifdim\dimen0>\dimen1                        % #1 is bigger
      \rlap{\hbox to \dimen0{\hfil/\hfil}}      % so center / in box
      #1                                        % and print #1
   \else                                        % / is bigger
      \rlap{\hbox to \dimen1{\hfil$#1$\hfil}}   % so center #1
      /                                         % and print /
   \fi}

\def\foursqr#1#2{{\vcenter{\vbox{
    \hrule height.#2pt
    \hbox{\vrule width.#2pt height#1pt \kern#1pt
    \vrule width.#2pt}
    \hrule height.#2pt
    \hrule height.#2pt
    \hbox{\vrule width.#2pt height#1pt \kern#1pt
    \vrule width.#2pt}
    \hrule height.#2pt
        \hrule height.#2pt
    \hbox{\vrule width.#2pt height#1pt \kern#1pt
    \vrule width.#2pt}
    \hrule height.#2pt
        \hrule height.#2pt
    \hbox{\vrule width.#2pt height#1pt \kern#1pt
    \vrule width.#2pt}
    \hrule height.#2pt}}}}
\def\psqr#1#2{{\vcenter{\vbox{\hrule height.#2pt
    \hbox{\vrule width.#2pt height#1pt \kern#1pt
    \vrule width.#2pt}
    \hrule height.#2pt \hrule height.#2pt
    \hbox{\vrule width.#2pt height#1pt \kern#1pt
    \vrule width.#2pt}
    \hrule height.#2pt}}}}
\def\sqr#1#2{{\vcenter{\vbox{\hrule height.#2pt
    \hbox{\vrule width.#2pt height#1pt \kern#1pt
    \vrule width.#2pt}
    \hrule height.#2pt}}}}

%\draftmode

%\cite{Eden}
\lref\Eden{
  R.~J.~Eden, P.~V.~Landshoff, D.~I.~Olive, J.~C.~Polkinghorne,
   {\it The Analytic S-Matrix,} Cambridge University Press (2002).
  %Phys.\ Rev.\ Lett.\  {\bf 41}, 1585 (1978).
  %%CITATION = PRLTA,41,1585;%%
  %254 citations counted in INSPIRE as of 29 Mar 2013
}
%\AdamsSV
\lref\AdamsSV{
  A.~Adams, N.~Arkani-Hamed, S.~Dubovsky, A.~Nicolis and R.~Rattazzi,
  %``Causality, analyticity and an IR obstruction to UV completion,''
JHEP {\bf 0610}, 014 (2006).
[hep-th/0602178].
%%CITATION = hep-th/0602178%%
}

%\VafaUI
\lref\VafaUI{
  C.~Vafa,
  %``The String landscape and the swampland,''
[hep-th/0509212].
%%CITATION = hep-th/0509212%%
}

%\ArkaniHamedDZ
\lref\ArkaniHamedDZ{
  N.~Arkani-Hamed, L.~Motl, A.~Nicolis and C.~Vafa,
  %``The String landscape, black holes and gravity as the weakest force,''
JHEP {\bf 0706}, 060 (2007).
[hep-th/0601001].
%%CITATION = hep-th/0601001%%
}

%\MaldacenaRE
\lref\MaldacenaRE{
  J.~M.~Maldacena,
  %``The Large N limit of superconformal field theories and supergravity,''
Adv.\ Theor.\ Math.\ Phys.\  {\bf 2}, 231 (1998).
[hep-th/9711200].
%%CITATION = hep-th/9711200%%
}

%\GubserBC
\lref\GubserBC{
  S.~S.~Gubser, I.~R.~Klebanov and A.~M.~Polyakov,
  %``Gauge theory correlators from noncritical string theory,''
Phys.\ Lett.\ B {\bf 428}, 105 (1998).
[hep-th/9802109].
%%CITATION = hep-th/9802109%%
}

%\WittenQJ
\lref\WittenQJ{
  E.~Witten,
  %``Anti-de Sitter space and holography,''
Adv.\ Theor.\ Math.\ Phys.\  {\bf 2}, 253 (1998).
[hep-th/9802150].
%%CITATION = hep-th/9802150%%
}

%\BriganteGZ
\lref\BriganteGZ{
  M.~Brigante, H.~Liu, R.~C.~Myers, S.~Shenker and S.~Yaida,
  %``The Viscosity Bound and Causality Violation,''
Phys.\ Rev.\ Lett.\  {\bf 100}, 191601 (2008).
[arXiv:0802.3318 [hep-th]].
%%CITATION = arXiv:0802.3318%%
}

%\HofmanUG
\lref\HofmanUG{
  D.~M.~Hofman,
  %``Higher Derivative Gravity, Causality and Positivity of Energy in a UV complete QFT,''
Nucl.\ Phys.\ B {\bf 823}, 174 (2009).
[arXiv:0907.1625 [hep-th]].
%%CITATION = arXiv:0907.1625%%
}

%\CamanhoHU
\lref\CamanhoHU{
  X.~O.~Camanho and J.~D.~Edelstein,
  %``Causality in AdS/CFT and Lovelock theory,''
JHEP {\bf 1006}, 099 (2010).
[arXiv:0912.1944 [hep-th]].
%%CITATION = arXiv:0912.1944%%
}

%\HofmanAR
\lref\HofmanAR{
  D.~M.~Hofman and J.~Maldacena,
  %``Conformal collider physics: Energy and charge correlations,''
JHEP {\bf 0805}, 012 (2008).
[arXiv:0803.1467 [hep-th]].
%%CITATION = arXiv:0803.1467%%
}

%\BuchelSK
\lref\BuchelSK{
  A.~Buchel, J.~Escobedo, R.~C.~Myers, M.~F.~Paulos, A.~Sinha and M.~Smolkin,
  %``Holographic GB gravity in arbitrary dimensions,''
JHEP {\bf 1003}, 111 (2010).
[arXiv:0911.4257 [hep-th]].
%%CITATION = arXiv:0911.4257%%
}

%\KulaxiziJT
\lref\KulaxiziJT{
  M.~Kulaxizi and A.~Parnachev,
  %``Energy Flux Positivity and Unitarity in CFTs,''
Phys.\ Rev.\ Lett.\  {\bf 106}, 011601 (2011).
[arXiv:1007.0553 [hep-th]].
%%CITATION = arXiv:1007.0553%%
}

%\BashamIQ
\lref\BashamIQ{
  C.~L.~Basham, L.~S.~Brown, S.~D.~Ellis and S.~T.~Love,
  %``Electron - Positron Annihilation Energy Pattern in Quantum Chromodynamics: Asymptotically Free Perturbation Theory,''
Phys.\ Rev.\ D {\bf 17}, 2298 (1978)..
%%CITATION = RLO-1388-746%%
}

%\BashamBW
\lref\BashamBW{
  C.~L.~Basham, L.~S.~Brown, S.~D.~Ellis and S.~T.~Love,
  %``Energy Correlations in electron - Positron Annihilation: Testing QCD,''
Phys.\ Rev.\ Lett.\  {\bf 41}, 1585 (1978)..
%%CITATION = RLO-1388-759%%
}

%\BashamGH
\lref\BashamGH{
  C.~L.~Basham, L.~S.~Brown, S.~D.~Ellis and S.~T.~Love,
  %``Energy Correlations In Perturbative Quantum Chromodynamics: A Conjecture For All Orders,''
Phys.\ Lett.\ B {\bf 85}, 297 (1979)..
%%CITATION = RLO-1388-786%%
}

%\DokshitzerSH
\lref\DokshitzerSH{
  Y.~L.~Dokshitzer, G.~Marchesini and B.~R.~Webber,
  %``Nonperturbative effects in the energy energy correlation,''
JHEP {\bf 9907}, 012 (1999).
[hep-ph/9905339].
%%CITATION = hep-ph/9905339%%
}

%\BrockSZ
\lref\BrockSZ{
  R.~Brock {\it et al.}  [CTEQ Collaboration],
  %``Handbook of perturbative QCD: Version 1.0,''
Rev.\ Mod.\ Phys.\  {\bf 67}, 157 (1995)..
%%CITATION = FERMILAB-PUB-93-094%%
}

%\SveshnikovVI
\lref\SveshnikovVI{
  N.~A.~Sveshnikov and F.~V.~Tkachov,
  %``Jets and quantum field theory,''
Phys.\ Lett.\ B {\bf 382}, 403 (1996).
[hep-ph/9512370].
%%CITATION = hep-ph/9512370%%
}

%\KorchemskySY
\lref\KorchemskySY{
  G.~P.~Korchemsky, G.~Oderda and G.~F.~Sterman,
  %``Power corrections and nonlocal operators,''
[hep-ph/9708346].
%%CITATION = hep-ph/9708346%%
}

%\CherzorAK
\lref\CherzorAK{
  P.~S.~Cherzor and N.~A.~Sveshnikov,
  %``Jet observables and energy momentum tensor,''
[hep-ph/9710349].
%%CITATION = hep-ph/9710349%%
}

%\KorchemskyKT
\lref\KorchemskyKT{
  G.~P.~Korchemsky and G.~F.~Sterman,
  %``Power corrections to event shapes and factorization,''
Nucl.\ Phys.\ B {\bf 555}, 335 (1999).
[hep-ph/9902341].
%%CITATION = hep-ph/9902341%%
}

%\BraunRP
\lref\BraunRP{
  V.~M.~Braun, G.~P.~Korchemsky and D.~Mueller,
  %``The Uses of conformal symmetry in QCD,''
Prog.\ Part.\ Nucl.\ Phys.\  {\bf 51}, 311 (2003).
[hep-ph/0306057].
%%CITATION = hep-ph/0306057%%
}

%\GrinsteinQK
\lref\GrinsteinQK{
  B.~Grinstein, K.~A.~Intriligator and I.~Z.~Rothstein,
  %``Comments on Unparticles,''
Phys.\ Lett.\ B {\bf 662}, 367 (2008).
[arXiv:0801.1140 [hep-ph]].
%%CITATION = arXiv:0801.1140%%
}

\lref\landauten{
L.~D.~Landau and E.~M.~Lifshitz, {\it Physical Kinetics,} Pergamon,
New York (1981).
}

\lref\ourtoappear{
A.~V.~Belitsky, S.~Hohenegger, G.~P.~Korchemsky, E.~Sokatchev, A.~Zhiboedov, {\it to appear}
}

%\MackJE
\lref\MackJE{
  G.~Mack,
  %``All Unitary Ray Representations of the Conformal Group SU(2,2) with Positive Energy,''
Commun.\ Math.\ Phys.\  {\bf 55}, 1 (1977)..
%%CITATION = DESY 75/50%%
}

%\KomargodskiVJ
\lref\KomargodskiVJ{
  Z.~Komargodski and A.~Schwimmer,
  %``On Renormalization Group Flows in Four Dimensions,''
JHEP {\bf 1112}, 099 (2011).
[arXiv:1107.3987 [hep-th]].
%%CITATION = arXiv:1107.3987%%
}

%\LutyWW
\lref\LutyWW{
  M.~A.~Luty, J.~Polchinski and R.~Rattazzi,
  %``The $a$-theorem and the Asymptotics of 4D Quantum Field Theory,''
JHEP {\bf 1301}, 152 (2013).
[arXiv:1204.5221 [hep-th]].
%%CITATION = arXiv:1204.5221%%
}

%\FitzpatrickYX
\lref\FitzpatrickYX{
  A.~L.~Fitzpatrick, J.~Kaplan, D.~Poland and D.~Simmons-Duffin,
  %``The Analytic Bootstrap and AdS Superhorizon Locality,''
[arXiv:1212.3616 [hep-th]].
%%CITATION = arXiv:1212.3616%%
}

%\KomargodskiEK
\lref\KomargodskiEK{
  Z.~Komargodski and A.~Zhiboedov,
  %``Convexity and Liberation at Large Spin,''
[arXiv:1212.4103 [hep-th]].
%%CITATION = arXiv:1212.4103%%
}

%\GiombiMS
\lref\GiombiMS{
  S.~Giombi and X.~Yin,
  %``The Higher Spin/Vector Model Duality,''
[arXiv:1208.4036 [hep-th]].
%%CITATION = arXiv:1208.4036%%
}

%\RattazziPE
\lref\RattazziPE{
  R.~Rattazzi, V.~S.~Rychkov, E.~Tonni and A.~Vichi,
  %``Bounding scalar operator dimensions in 4D CFT,''
JHEP {\bf 0812}, 031 (2008).
[arXiv:0807.0004 [hep-th]].
%%CITATION = arXiv:0807.0004%%
}

%\ElShowkHT
\lref\ElShowkHT{
  S.~El-Showk, M.~F.~Paulos, D.~Poland, S.~Rychkov, D.~Simmons-Duffin and A.~Vichi,
  %``Solving the 3D Ising Model with the Conformal Bootstrap,''
Phys.\ Rev.\ D {\bf 86}, 025022 (2012).
[arXiv:1203.6064 [hep-th]].
%%CITATION = arXiv:1203.6064%%
}

%\BeemQXA
\lref\BeemQXA{
  C.~Beem, L.~Rastelli and B.~C.~van Rees,
  %``The N=4 Superconformal Bootstrap,''
[arXiv:1304.1803 [hep-th]].
%%CITATION = YITP-SB-13-10%%
}

\lref\Istrail{
S.~Istrail, {\it Proceedings of the 32nd ACM
  Symposium on the Theory of Computing (STOC00), ACM Press, Portland, Oregon,
  May 21-23, 2000}, pp.~87--96.
}
%``{Statistical Mechanics, Three-Dimensionality and NP-Completeness:
  %I. Universality of Intractability of the Partition Functions of the Ising
  %Model Across Non-Planar Lattices},'

\lref\dotspol{
Vl.~S.~Dotsenko, A.~M.~Polyakov, %Fermion representations for the 2D and 3D Ising models,
Adv. Stud. Pure Math.\ {\bf 16}, 171-203 (1988)
}

%\DistlerRR
\lref\DistlerRR{
  J.~Distler,
  %``A Note on the 3-D Ising model as a string theory,''
Nucl.\ Phys.\ B {\bf 388}, 648 (1992).
[hep-th/9205100].
%%CITATION = hep-th/9205100%%
}
%\OsbornCR
\lref\OsbornCR{
  H.~Osborn and A.~C.~Petkou,
  %``Implications of conformal invariance in field theories for general dimensions,''
Annals Phys.\  {\bf 231}, 311 (1994).
[hep-th/9307010].
%%CITATION = hep-th/9307010%%
}

%\ErdmengerYC
\lref\ErdmengerYC{
  J.~Erdmenger and H.~Osborn,
  %``Conserved currents and the energy momentum tensor in conformally invariant theories for general dimensions,''
Nucl.\ Phys.\ B {\bf 483}, 431 (1997).
[hep-th/9605009].
%%CITATION = hep-th/9605009%%
}

%\Dymarsky
\lref\Dymarsky{
  A.~Dymarsky,
  %``Conserved currents and the energy momentum tensor in conformally invariant theories for general dimensions,''
 {\it On the Four-point Function of the Stress-energy Tensor in a CFT, to appear}
  %%CITATION = HEP-TH/9605009;%%
  %104 citations counted in INSPIRE as of 21 Mar 2013
}

%\LiendoHY
\lref\LiendoHY{
  P.~Liendo, L.~Rastelli and B.~C.~van Rees,
  %``The Bootstrap Program for Boundary CFT_d,''
[arXiv:1210.4258 [hep-th]].
%%CITATION = arXiv:1210.4258%%
}

\lref\gelfandshilov{
I.~Gelfand, G.~Shilov, ``Generalized functions,'' Acad.  Press, New York-London (1964).
}

\lref\vladimirov{
V.~S.~Vladimirov,  ``Generalized functions in mathematical physics,'' p. 52, Moscow Izdatel Nauka (1976)
}

%\WeinbergKQ
\lref\WeinbergKQ{
  S.~Weinberg and E.~Witten,
  %``Limits on Massless Particles,''
Phys.\ Lett.\ B {\bf 96}, 59 (1980)..
%%CITATION = HUTP-80/A056%%
}

%\BauerDT
\lref\BauerDT{
  C.~W.~Bauer, S.~P.~Fleming, C.~Lee and G.~F.~Sterman,
  %``Factorization of e+e- Event Shape Distributions with Hadronic Final States in Soft Collinear Effective Theory,''
Phys.\ Rev.\ D {\bf 78}, 034027 (2008).
[arXiv:0801.4569 [hep-ph]].
%%CITATION = arXiv:0801.4569%%
}

%\CornalbaFS
\lref\CornalbaFS{
  L.~Cornalba,
  %``Eikonal methods in AdS/CFT: Regge theory and multi-reggeon exchange,''
[arXiv:0710.5480 [hep-th]].
%%CITATION = arXiv:0710.5480%%
}

%\cite{Gantmaher}
\lref\Gantmaher{
  F.~R.~Gantmacher,
  {\it The Theory of Matrices, Vol. 1,}
  p.307, AMS (1990)
  %%CITATION = ARXIV:1210.4258;%%
  %4 citations counted in INSPIRE as of 21 Mar 2013
}

%\ZhiboedovBM
\lref\ZhiboedovBM{
  A.~Zhiboedov,
  %``A note on three-point functions of conserved currents,''
[arXiv:1206.6370 [hep-th]].
%%CITATION = arXiv:1206.6370%%
}

%\CostaMG
\lref\CostaMG{
  M.~S.~Costa, J.~Penedones, D.~Poland and S.~Rychkov,
  %``Spinning Conformal Correlators,''
JHEP {\bf 1111}, 071 (2011).
[arXiv:1107.3554 [hep-th]].
%%CITATION = arXiv:1107.3554%%
}

%\MaldacenaJN
\lref\MaldacenaJN{
  J.~Maldacena and A.~Zhiboedov,
  %``Constraining Conformal Field Theories with A Higher Spin Symmetry,''
[arXiv:1112.1016 [hep-th]].
%%CITATION = arXiv:1112.1016%%
}

%\StermanWJ
\lref\StermanWJ{
  G.~F.~Sterman and S.~Weinberg,
  %``Jets from Quantum Chromodynamics,''
Phys.\ Rev.\ Lett.\  {\bf 39}, 1436 (1977)..
%%CITATION = HUTP-77/A044%%
}

%\HawkingSW
\lref\HawkingSW{
  S.~W.~Hawking,
  %``Particle Creation by Black Holes,''
Commun.\ Math.\ Phys.\  {\bf 43}, 199 (1975), [Erratum-ibid.\  {\bf 46}, 206 (1976)]..
}

\rightline{PUPT-2444}
\Title{
} {\vbox{\centerline{On Conformal Field Theories With Extremal ${a \over c}$ Values}
\vskip3pt
%\centerline{}
}}

\centerline{Alexander Zhiboedov}
\bigskip

\centerline{{\it Department of Physics, Princeton University, Princeton, NJ 08544, USA}}

\vskip20pt

Unitary conformal field theories (CFTs) are believed to have positive (non-negative) energy correlators. Energy correlators are universal observables in higher-dimensional CFTs built out of integrated Wightman functions of the stress-energy tensor. We analyze energy correlators in parity invariant four-dimensional CFTs. The goal is to use
the positivity of energy correlators to further constrain unitary CFTs. It is known that the positivity of the simplest one-point energy correlator implies that ${1 \over 3} \leq {a \over c} \leq {31 \over 18}$ where $a$ and $c$ are the Weyl anomaly coefficients. We use the {\it positivity} of higher point energy correlators to show that CFTs with extremal values of ${a \over c}$ have trivial scattering observables. More precisely, for ${a \over c} = {1 \over 3}$ and ${a \over c} = {31 \over 18}$ all energy correlators are fixed to be the ones of the free boson and the free vector theory correspondingly.  Similarly, we show that the {\it positivity} and {\it finiteness} of energy correlators together imply that the three-point function of the stress tensor in a CFT cannot be proportional to the one in the theory of free boson, free fermion or free vector field.

\noindent

\vskip 7pt

\Date{April 2013}

%\listtoc \writetoc
%\vskip 0.5in

%\draftmode

\newsec{Introduction}

Restrictions imposed by causality and unitarity on the S-matrix were understood a long time ago \Eden. Similarly, in conformal
field theories unitarity is known to impose constraints on scaling dimensions of operators and their two-point functions \MackJE~
traditionally derived from reflection positivity. Alternatively, unitarity bounds arise from the positivity of the total cross section \GrinsteinQK~ which is defined through the Fourier transform of the Wightman two-point function.

The notions of a scattering amplitude and differential cross section do not exist in a generic CFT. In accordance with this, there is no analog of the differential cross section positivity. However, close relatives of differential cross sections that do exist in any CFT are known. These are so-called weighted cross sections \BrockSZ~which are more inclusive than the differential cross section but less inclusive than the total one.

Energy correlators are a famous exemplar of the weighted cross section type of observable that is positively definite. They first appeared in the study of electron-positron annihilation in QCD \refs{\BashamBW,\BashamGH}.

\ifig\setup{The experimental setup. The red region corresponds to a vacuum excitation produced in our case by an insertion of a local operator. Blue points stand for detectors that collect energy at infinity.} {\epsfxsize1.2in\epsfbox{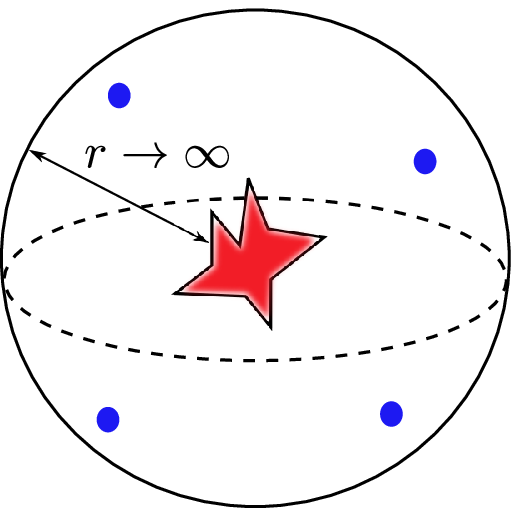}}

The following (see \setup) is a physical picture behind these observables. We hit the vacuum with a local operator, let the excitations evolve and study the energy distribution at infinity. We repeat the experiment many times and compute the mean. The result of such experiment provides the experimentalist with many hints about the underlying physical dynamics. Probably, the most famous energy distributions are the QCD jets \StermanWJ\ and Hawking radiation \HawkingSW. The $n$-point energy correlator corresponds to probing the state with $n$ calorimeters. In theories with massless particles energy correlators are known to be infrared safe or finite.  Positivity of the differential cross section together with stability of the vacuum imply positivity of energy correlators in theories with the S-matrix.

Another important property of energy correlators is that they could be reformulated in terms of correlation functions \refs{\SveshnikovVI,\KorchemskySY,\CherzorAK,\KorchemskyKT} so that the notion of the S-matrix is not necessary to define them. IR-safety and existence of the operator formulation suggest that energy correlators should be well-defined in CFTs as well, indeed, it is the case \HofmanAR.  The $n$-point energy correlator is defined starting from the $(n+2)$-point Wightman function with $n$ insertions of the stress tensor corresponding to calorimeters and two Hermitian conjugate operators corresponding to the nontrivial state under consideration. As in theories with a unitary S-matrix one can expect that energy correlators in unitary CFTs should be positive. There are some plausible arguments supporting this claim in the literature \refs{\HofmanAR,\HofmanUG,\KulaxiziJT}; however, a general proof is still lacking.

A Euclidean analog of energy correlators positivity is not known. It is similar to many instances when it is not clear how to reproduce positivity constraints that are trivial in Minkowski space from Euclidean correlation functions (for recent examples see \refs{\KomargodskiVJ,\LutyWW,\FitzpatrickYX,\KomargodskiEK}). Interestingly, through AdS/CFT \refs{\MaldacenaRE,\GubserBC,\WittenQJ} the positivity of energy correlators is related to causality in AdS space \refs{\BriganteGZ,\HofmanUG,\CamanhoHU}.\foot{See also \refs{\AdamsSV,\VafaUI,\ArkaniHamedDZ}.}

By assuming energy correlators to be positive what kind of constraints do we get? The case of a one-point energy correlator was analyzed before \refs{\HofmanAR,\BuchelSK}. We will review this constraint below but for now let us write its consequence
\eqn\ratioac{
{1 \over 3} \leq {a \over c} \leq {31 \over 18}
}
where $a$ and $c$ are the Weyl anomaly coefficients.

In this note we analyze the positivity constraint for higher point energy correlators in parity invariant 4d CFTs. In particular, we show that for ${a \over c} = {1 \over 3}$ and ${a \over c} = {31 \over 18}$ the positivity of energy correlators imply their triviality. Namely they are the ones of the free boson and free vector field correspondingly.

The energy correlators in free theories are not functions but distributions since they contain terms like\foot{$\theta$ stands for the angle between detectors.} $\delta(\theta - \pi)$ that come from two free particles propagating in opposite directions. In interacting theories such singular terms are believed to be absent due to the copious production of soft particles so that energy correlators are finite at non-coincident points. We show that under the additional assumption of finiteness of energy correlators at non-coincident points it is possible to show that the three-point function of the stress tensor in a CFT cannot be proportional to the one in the theory of free boson, free fermion or free vector field. The assumption of finiteness is consistent with all known results for energy correlators in interacting CFTs.

Let us emphasize that in this note we do not assume anything about the spectrum of operators in the theory apart from the existence of the stress-energy tensor. All arguments are based solely on the positivity of energy correlators and conformal symmetry and, thus, are non-perturbative and applicable to all known CFTs.

\newsec{Basics of Energy Correlators}

In this section we explicate the definition and basic properties of energy correlators. Besides kinematics these include positivity, a total flux bound and a zero flux constraint.

\subsec{Three-point Function of the Stress Tensor}

Before discussing energy correlators let us review some basic facts about the stress tensor correlation functions in CFTs.
Conformal symmetry fixes the three-point function of the stress tensor
in terms of three real numbers \refs{\OsbornCR,\ErdmengerYC}.\foot{The two-point function is fixed up to one number.} It is convenient to choose these numbers as follows. Consider two and three-point functions in the theory of free boson,
free Dirac fermion and free vector fields. Let us denote them $\la T T\ra_{{\rm b}}$,$\la T T T\ra_{{\rm b}}$ and similarly for the fermion and vector structures. The stress tensor two- and three- point functions take the form
\eqn\twoandthree{\eqalign{
\la T T\ra &= n_b \la T T\ra_{{\rm b}} + n_f \la T T\ra_{{\rm f}} + n_v \la T T\ra_{{\rm v}}, \cr
\la T T T \ra &= n_b \la T T T\ra_{{\rm b}} + n_f \la T T T\ra_{{\rm f}} + n_v \la T T T \ra_{{\rm v}}.
}}
In this way for every CFT we can talk about an effective number of bosons, fermions and vectors. We would like to stress that this labeling has nothing to do with the theory content and is a purely bookkeeping tool.  Reflection positivity of the two-point function of the stress tensor enforces
\eqn\twoppos{
{1 \over 3} n_b + 2 n_{f} + 4 n_v >0.
}
Starting from four points correlation functions depend on arbitrary functions of conformal cross ratios \refs{\CostaMG, \Dymarsky}.

\subsec{Definition of Energy Correlators}

Energy correlators can be defined using a detector operator built out of the stress tensor
\refs{\SveshnikovVI,\KorchemskySY,\CherzorAK,\KorchemskyKT,\HofmanAR}
\eqn\detA{
{\cal E}(\vec n) = \lim_{r \to \infty} r^2 \int_{- \infty}^{\infty} d t T_{0 i} n^i (t, r \vec n).
}
The integral above is the detector working time integral, a period of time during which the detector is collecting energy. The limit appears because we place the detector at infinity far away from the ``collision'' region. Thanks to the insertion point being $r \vec n$ the calorimeter collects only the radiation propagating in the $\vec n$ direction.

In this note we define a detector operator in a slightly different manner (see also \ourtoappear)
\eqn\detB{
{\cal E}(n) =  \int_{-\infty}^{\infty} d (x .n)  \lim_{x . \bar n \to \infty} { (x . \bar n) ^2 \over 16} T.\bar n . \bar n (x .n,x .\bar n,\vec 0_{\perp})
}
where $n=(1, \vec n)$, $\bar n=(1,-\vec n)$ so that $x .n = t-r$, $x . \bar n = t+r$ and $\vec 0_{\perp}$ stands for coordinates perpendicular to $n$ and $\bar n$.\foot{Notations here come from the Sudakov decomposition of the four momentum $q = \alpha n + \beta \bar n + \vec q_{\perp}$ common in the QCD literature \BraunRP.}

\ifig\detectors{Penrose diagrams of Minkowski space. a)  The definition \detA~states that we first collect energy at all times for fixed distance and then take the distance to infinity. b) Definition \detB~is designed for CFTs. From the very beginning we sent the detector to the future null infinity ${\cal I}^{+}$ and the integrate over the null working time of the detector.} {\epsfxsize1.6in\epsfbox{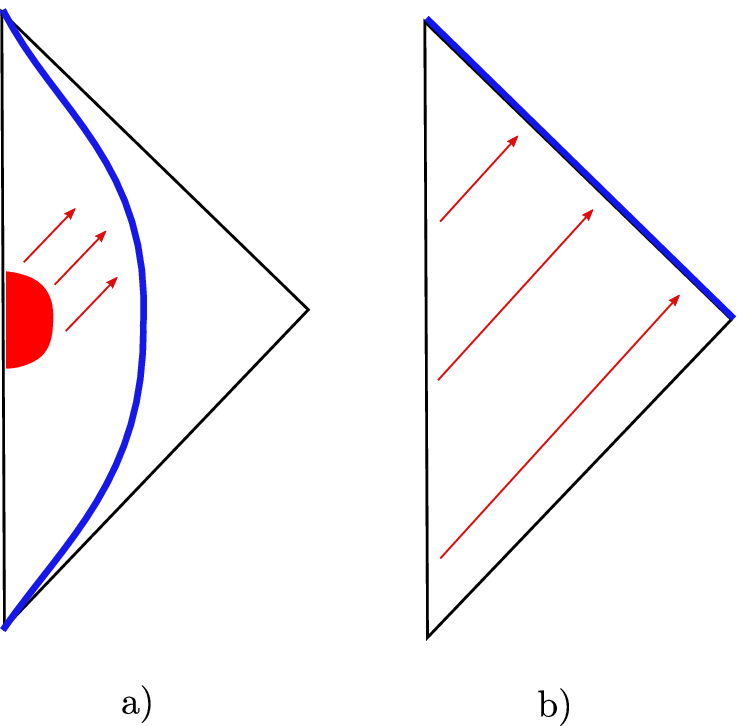}}

Notice that in general two definitions are not equivalent. Indeed, the definition \detA~ corresponds to the detector that collects all sorts of excitations, massive (propagating to the future and from the past time-like infinity $i^{\pm}$) as well as massless (propagating to and from the null infinity ${\cal I}^{\pm}$). A usual physical setup corresponds to the choice of the state $| \Psi \ra$ with a compact support, such that the detector triggers only the outgoing particles propagating to $i^+$ and ${\cal I}^{+}$. At the very end of the calculation we can take the limit in which the wave function can become a simple plane wave as we consider below. If we instead start from the wave function which is nowhere zero we will detect the ingoing particles that hit the detector from the outside. In CFTs for the simple states created by the primary operator with momentum $q$ it is easy to check that energy flux at $i^+$ is zero.

The definition \detB~ is designed to collect only radiation at ${\cal I}^+$ from the very beginning. We first send the detector to the future null infinity and then integrate over the null working time. We can use simple plane wave states since it is impossible to hit the detector inserted along the null line from the outside. Another advantage of \detB~ is that it is manifestly Lorentz invariant. Thus, qualitatively we see that in CFTs the two definitions are equivalent but the second one is a bit more convenient to handle in practice. The quantitative equivalence of the two definitions as well as the existence of the limit could be seen using the global conformal transformations explained in \HofmanAR. We review it in appendix A.

Energy correlators are defined as follows
\eqn\energydef{
\la {\cal E}(n_1) ... {\cal E}(n_k) \ra_{\Psi} = {\la \Psi | {\cal E}(n_1) ... {\cal E}(n_k) | \Psi \ra \over \la \Psi | \Psi \ra}.
}
In \energydef~ all correlation functions are Wightman or non-ordered functions. Non-ordered correlation functions are typical to the in-in type of computations that we discuss here \landauten. A universal\foot{This state always exists.} state we would like to consider is obtained by acting with the stress tensor carrying momentum $q$ on the vacuum\foot{Our sign convention is $(+---)$.}
\eqn\state{
| T.\eps.\eps(q) \ra = \int d^4 x e^{-i q x} T_{\mu \nu}(x) \eps^{\mu} \eps^{\nu} | 0 \ra,
}
where $\eps^{\mu}$ is the polarization tensor.\foot{As it is written $\la \Psi | \Psi \ra$ is infinite. This is not a problem. We can consider Gaussian wave packets instead to make the norm of the state finite or, more conveniently, put the in- stress tensor at the origin in the coordinate space so that $\la \Psi | \Psi \ra = \int d^4 x e^{i q x} \la 0 | T.\eps^*.\eps^*(x) T.\eps.\eps(0)| 0 \ra < \infty$. This is what we do in this note.} Since the stress tensor is conserved, symmetric and traceless we choose $\eps^2 = 0$, $\eps .q  = 0$.
The momentum $q$ is always assumed to be time-like with positive energy $q^0 >0$.

Throughout the paper we freely switch between the polarization tensor being $\eps^{\mu} \eps^{\nu}$ with $\eps^2 = 0$, $\eps.q = 0$ and $\eps^{\mu \nu}$ with $\eps^{\mu}_{~ \mu} = 0$, $q_{\mu}\eps^{\mu \nu} = 0$. We found it more convenient to use the first choice when writing the most general structures allowed by symmetries while the second is more convenient when thinking about the positivity of energy correlators in the reference frame where $q=(q^0, \vec 0)$ as will be explained below.

\subsec{Symmetries}

Let us discuss symmetries of energy correlators with the state \state. They are

\item{a)} Lorentz invariance
\eqn\lorentz{
\la {\cal E}(\Lambda n_1) ... {\cal E}(\Lambda n_k) \ra_{T.\Lambda\eps.\Lambda \eps(\Lambda q)} = \la {\cal E}(n_1) ... {\cal E}(n_k) \ra_{T.\eps.\eps(q)} .
}

This is manifest in the way we defined energy correlators above \detB.

\item{b)} Projective covariance
\eqn\proj{
\la {\cal E}(\lambda n_1) ... {\cal E}(n_k) \ra_{T.\eps.\eps(q)} = \lambda^{-3} \la {\cal E}(n_1) ... {\cal E}(n_k) \ra_{T.\eps.\eps(q)} .
}
In the formula above we relaxed the condition $n=(1,\vec n)$ and considered $n$ to be a light-like vector $n^2 = 0$.
The property \proj\ could be easily checked to follow from conformal properties of the stress tensor.
Qualitatively, one can understand it as follows. The leading contribution from the spin part of $T(x).\bar n . \bar n$ behaves like $(x . \bar n)^2$.
Combined with the limiting prefactor this forms $(x . \bar n)^4$ that picks up the corresponding term in the large $(x . \bar n)$ expansion of the correlator. By Lorentz invariance the expansion takes the schematic form ${1 \over (x-y. \bar n)^4 (x -y. n)^4}$ which scales like $\lambda^{-4}$ under rescaling of $n$. The additional factor of $\lambda$ comes from the measure $\int_{- \infty}^{\infty} d (x .n)$.

\item{c)} Momentum rescaling
\eqn\momentum{
\la {\cal E}(n_1) ... {\cal E}(n_k) \ra_{T.\eps.\eps(\lambda q)} = \lambda^{k} \la {\cal E}(n_1) ... {\cal E}(n_k) \ra_{T.\eps.\eps(q)}.
}
This follows from the fact that each detector measures energy. Equivalently, it could be seen using dilatation symmetry.

\item{d)} Permutation symmetry $n_{i} \leftrightarrow n_j$. Detectors are space-like separated and therefore commute.

\item{e)} Reality. Reality of energy correlators follows from Hermiticity of the stress-energy tensor.

More details and concrete examples of energy correlator computations can be found in \ourtoappear. It is clear that we can think about energy correlators as correlation functions of scalar primary operators in 2d CFT with a vacuum that breaks conformal symmetry  \HofmanAR. If we choose the state to be given by a scalar operator instead of a tensor then kinematics of energy correlators is identical to that of BCFT \LiendoHY\ where the role of the boundary is played by the momentum of the operator. This analogy could be useful when thinking about the small angle OPE for detector operators since instead of thinking about the light-cone OPE of light-ray operators we are dealing with the usual OPE in 2d Euclidean CFT even though the structure of spectrum in this case is different.

\subsec{One- and Two-point Energy Correlators}

The one-point energy correlator originates from the three-point function of the stress tensor which itself depends on three real numbers as was reviewed above.  This fact together with $\int d \Omega_{n} \la {\cal E}(n)\ra_{T.\eps.\eps(q)} = \la H \ra_{T.\eps.\eps(q)} = q^0$ where $H$ stands for the Hamiltonian operator fix the energy correlator up to two numbers \refs{\HofmanAR,\BuchelSK}
\eqn\onepointpos{\eqalign{
\la {\cal E}(n)\ra_{T.\eps.\eps(q)} = {q^4 \over 4 \pi (q.n)^3} \left(1 + t_2 \left[ {q^2 \eps.n \eps^*.n \over (q.n)^2 \eps.\eps^*} - {1 \over 3} \right] +t_4 \left[ {q^4 \eps.n^2 \eps^*.n^2 \over (q.n)^4(\eps. \eps^*)^2} - {2 \over 15} \right] \right).
}}
The relation between $t_{2,4}$ and $n_{b,f,v}$ defined in section 2.1 can be found in appendix C of \HofmanAR.\foot{$n_f^{here} = {1 \over 2} n_{wf}^{there}$.}

The two-point energy correlator is built out of the four-point function of the stress tensor.
Its most general form is very complicated \Dymarsky. Fortunately, the two-point energy correlator built out of it is much simpler. The freedom that we have in this case is still functional because we can have arbitrary functions of the cross ratio $\xi = {q^2 n_1.n_2 \over 2 q.n_1 q.n_2}$.\foot{Notice that $0 < \xi \leq 1$ for time-like $q$. By excluding zero we avoid the situation when detectors are on top of each other.  From \detA~and \detB~it is clear that when this happens the stress tensor operators become light-like separated and it requires extra care to define energy correlators.} After imposing symmetry constraints we get for parity invariant theories
\eqn\twopoint{
\la {\cal E}(n_1)  {\cal E}(n_2) \ra_{T.\eps.\eps(q)} = {q^2 \over (n_1 . n_2)^3} \sum_{i=1}^{7} {\rm T}_{i} f_{i} (\xi),
}
where ${\rm T}_{i}(\eps,\eps^*,n_i,q)$ stand for possible tensor structures. Their precise form can be found in appendix B. In principle there are also parity violating structures that contain the epsilon tensor. We do not consider them in this note. The n-point function will depend on the cross ratios $\xi_{i j}=  {q^2 n_i.n_j \over 2 q.n_i q.n_j}$.

\subsec{Finiteness}

At this point we should ask ourselves, what type of functions $f_i(\xi)$ we are allowed to have in \twopoint ? As we mentioned in the introduction, already in free theories energy correlators contain terms of the type $\delta(1-\xi)$. Thus, in general we expect energy correlators to be functionals rather than functions. This becomes relevant starting from the two-point energy correlator since the one-point energy correlator is fixed by symmetries and is finite \onepointpos . In this section we summarize our assumptions about  energy correlators.

First, we imagine that real physical detectors have a finite angular resolution, thus, what we really measure is the integrated version of the energy correlator. So after smearing over angles we expect to get a finite result
\eqn\finiteness{
\la {\cal E}(n_1)  {\cal E}(n_2) \ra_{T.\eps.\eps(q)}[g] = \int_{\delta \Omega} d \Omega_{n_2} g(n_2)\la {\cal E}(n_1)  {\cal E}(n_2) \ra_{T.\eps.\eps(q)} < \infty
}
where we integrated over an arbitrary region on the sphere $\delta \Omega$ that does not contain the point $n_1$ and the measurement function $g \in C^{\infty}(S^2)$. We assume that energy correlators are linear functionals. Physically, it means that increasing the detector sensitivity by a factor of two at each point $g \to 2 g$ results the measurement to be twice as large as well. For the linear functionals we also assume the smoothness condition \gelfandshilov~that enters in the standard definition of a distribution.

More importantly, we assume that energy correlators are finite (regular) for non-coincident detector positions up to a set of {\it isolated} points $\{ \xi^*_i \}$ for which we can have singularities (irregularities) that integrate to a finite number \finiteness. In other words, we think about energy correlators as if they are regular functions up to a set of isolated points $\{ \xi^*_i \}$ where they are defined only in the functional sense.

Even though the terms like $\delta(1-\xi)$ do appear in free theories and in perturbation theory, available strong coupling data \HofmanAR\ as well as resummed perturbation theory analysis in QCD \refs{\BashamGH,\DokshitzerSH} show that delta functions disappear when all orders are included. So, it is reasonable to assume that in interacting CFTs energy correlators are finite when detectors are at non-coincident points. This is an assumption of {\it finiteness}. We will not use it unless stated otherwise.

\subsec{Positivity}

There is good evidence \refs{\HofmanAR,\HofmanUG,\KulaxiziJT} that energy correlators in unitary CFTs are non-negative
\eqn\positivity{
\la {\cal E}( n_1) ... {\cal E}(n_k) \ra_{\Psi} \geq 0
}
for any $\Psi$.\foot{We could have thought about the following argument. Let us assume that the CFT admits a deformation by a relevant operator $\delta {\cal L} = m^{d - \Delta} {\cal O}_{\Delta}(x)$ that drives it to the gapped phase. Since the theory is free at large distances the insertion of the stress tensor at infinity acts on the space of free particles producing the usual energy correlators defined in terms of scattering amplitudes \BauerDT. The fact that all particles carry positive energy together with the positivity of the cross section imply that $\la {\cal E}( n_1) ... {\cal E}(n_k) \ra_{\Psi} (m) \geq 0$ for any $m$. Energy correlators are known to be IR-safe or, in other words, this function is known to have a smooth limit $m \to 0$ which implies \positivity~if the following condition holds
\eqn\equalitySM{
\lim_{m \to 0} \la {\cal E}( n_1) ... {\cal E}(n_k) \ra_{\Psi} (m) = \la {\cal E}( n_1) ... {\cal E}(n_k) \ra_{\Psi}^{CFT}.
}
While the equalities of the type \equalitySM~are trivially true for local correlation functions when dealing with integrated correlation functions and their limits an issue of the order of limits can possibly arise. A trivial example is $\lim_{m \to 0} \lim_{r \to \infty} e^{- m r} = 0$, $\lim_{r \to \infty} \lim_{m \to 0} e^{- m r} = 1$. It will be interesting to understand when \equalitySM~holds.} This condition looks physically reasonable and holds in all known examples, though we are not aware of its general proof.

In light of the previous section we need to specify what do we exactly mean by \positivity. Indeed, for the configuration of detectors such that the energy correlator is finite or under assumption of finiteness we can use \positivity; however, in the case when the energy correlator is not regular we need to impose the positivity of the integrated energy correlator \finiteness~ with $g(\delta \Omega) \geq 0$ .

In \refs{\HofmanAR,\BuchelSK} it was shown that the positivity of the one-point function
\eqn\onepointpos{\eqalign{
\la {\cal E}(n)\ra_{T.\eps.\eps(q)} \geq 0
}}
implies the following inequalities for $n_{b,f,v}$\foot{Another solution is that all constants are non-positive. This contradicts the positivity of the two-point function \twoppos.}
\eqn\furtu{
n_b \geq 0, ~ n_f \geq 0, ~ n_v \geq 0.
}

A weaker version of this constraint was quoted in the introduction \ratioac,
where the value ${a \over c} = {1 \over 3}$ corresponds to $n_b \neq 0$,$ n_f = n_v =0$; the case ${a \over c} = {31 \over 18}$,
on the other hand, is given by $n_v \neq 0$,$ n_b = n_f =0$. One can ask if it is possible to derive more constraints from the positivity of higher point energy correlators?

Let us consider, for example, the two-point energy correlator. We can fix the position of one detector
and vary the position of the second detector and the polarization tensor such that the cross ratio $\xi$ is fixed. In this way the problem
is identical to the one of the one-point function. Indeed, $f_i (\xi)$ are just constants and we are interested in conditions
that the quadratic form $\eps^*_{i j} M^{i j k l} \eps_{k l}$ is non-negative for arbitrary $\eps$ and $n_2$ such that
$n_1.n_2$ is fixed. $ M^{i j k l}$ is a Hermitian matrix due to reality of energy correlators. A Hermitian matrix is non-negative if and only if all its eigenvalues are non-negative. In practice it is also useful to use another criterion (see e.g. \Gantmaher): a Hermitian matrix is non-negative if and only if all its principal minors are non-negative. If we choose the reference frame $q=(q^0, \vec 0)$ and $n_1 = (1,0,0,1)$ the positivity constraint takes the form
\eqn\positivityeigen{
\lambda_{M} \geq 0, ~ {\rm any} ~ (n_2^1)^2 + (n_2^2)^2 \leq 1 -  (n_2^3)^2 = 4 \xi (1-\xi)
}
where $\lambda_M$ are eigenvalues of $M$. Below we use the positivity constraint \positivityeigen~only in particular cases when it is especially simple.

\subsec{Total Flux Bound and Zero Flux Constraint}

The fact that we consider a state with energy $q^0$ implies the following inequality\foot{In this formula we implicitly assumed that $n^0_i = 1$. The only place in the paper where we relax this assumption is when we analyze symmetries and possible structures.}
\eqn\inegualv{\eqalign{
&\int_{\delta \Omega} d \Omega_{\vec n_i} \la  {\cal E}(n_1)...{\cal E}(n_i) ... {\cal E}( n_k) \ra_{\Psi} \leq q^0 \la {\cal E}(n_1) ... {\cal E}(n_k)  \ra_{\Psi},
}}
for any region on the sphere $\delta \Omega$ that does not include other detectors insertions. Physically, this condition is trivial. It states
that the energy calorimeter cannot detect energy larger than the total energy of the state. However, to prove it from the first principles is more tricky.
One way to proceed is to define energy correlators at coincident points. For example, one can think of taking the limits \detA,\detB\ for different detectors one after another such that detectors are always very far from each other. In this way the operator identity $\int d \Omega_{\vec n_i} {\cal E}(n_i) = q^0$ holds and imply \inegualv. We call \inegualv~{\it the total flux bound}.

One can check that something special happens when $n_b n_f n_v = 0$. In this case, for any direction $n$ it is possible to find a polarization tensor $\eps_{\mu \nu}$ such that
\eqn\zeroflux{
\la {\cal E}(n)\ra_{T^{\mu \nu} \eps_{\mu \nu}(q)} = 0.
}
This is not possible if $n_b n_f n_v > 0$. More generally, consider a situation when there exists $\hat n$ such that
\eqn\zerofluxB{
\la {\cal E}(\hat n)\ra_{\Psi} = 0.
}
It is clear that \positivity,\zerofluxB\ and \inegualv\ together imply that
\eqn\zeroen{
\la {\cal E}(n_1) ... {\cal E}( n_k)  {\cal E} (\hat n)  \ra_{\Psi} = 0.
}
Below we use this assertion to further constrain energy correlators and we call it {\it the zero flux constraint}.

\newsec{Argument}

In this section we use the positivity constraint \positivity~to fix energy correlators completely in the case of extremal values of ${a \over c}$. We also show that an additional assumption of energy correlators finiteness is not consistent with the three-point function of the stress tensor being proportional to that of the free boson, free fermion or free vector field.

First, we use the zero flux constraint \zeroen\ to reduce the number of possible functions in \twopoint. Second, we use the positivity constraint \positivity\ to fix energy correlators completely. The positivity constraint is used in two steps as well. We start by imposing the positivity constraint at the points where the energy correlator is finite. We then impose the integrated positivity constraint in the isolated singular points.

It is absolutely crucial for the argument that the stress tensor has nonzero spin.\foot{The famous example of an argument for which spin of the state is crucial was provided by Weinberg and Witten \WeinbergKQ .}
Imagine a state created by a scalar operator. The two-point energy correlator takes the form
\eqn\twopointSC{
\la {\cal E}(n_1)  {\cal E}(n_2) \ra_{{\cal O}(q)} = {q^2 \over (n_1 . n_2)^3}  f(\xi).
}
The positivity constraint states that $f(\xi) \geq 0$ and clearly is not strong enough to fix the energy correlator.

\subsec{Purely Bosonic Structure or ${a \over c} = {1 \over 3}$}

Let us assume that the three-point function of the stress tensor is given by the purely bosonic structure $n_b \neq 0$,
$n_f = 0$, $n_v = 0$. A priori nothing prevents us from thinking that there are interacting CFTs of this type. The one-point energy correlator takes the form

\eqn\onepointboson{
\la {\cal E}(\hat n) \ra_{T.\eps.\eps(q)} ={15 \over 8 \pi} {(q^2)^4 \over (q. \hat n)^7} {\eps. \hat n^2 \eps^*. \hat n^2 \over  (\eps . \eps^*)^2 }.
}
Henceforth it is more convenient to choose the reference frame $q = (q^0 , \vec 0)$ and $\hat n = (1,0,0,1)$ so we align the detector
along the $z$-axis. We also switch to the state $T_{i j} \eps^{i j}$ where $\eps^{i i} = 0$ and Latin indices run from $1$ to $3$.

As a first step we find polarization tensors such that the one-point function \onepointboson\ is zero. This condition imposes  $\eps_{3 3}=0$. Other than this the polarization tensor is arbitrary. We can use this polarization tensor to impose the zero flux constraint \zeroen\ on the two-point energy correlator
\eqn\zerofluxBB{
\la {\cal E}(n) {\cal E}(\hat n)  \ra_{T_{i j} \eps^{i j}(q^0)} = 0,~ {\rm any} ~ n.
}
Let us explain how we impose the zero flux constraint. We choose $\hat n$ along the $z$-axis and $\eps_{33} = 0$. The cross ratio becomes $\xi = {1 - n^3 \over 2}$ and we keep it fixed. This allows us to think of $f(\xi)$ as constant numbers. We consider then all possible polarization tensors and $n^{1,2}$ components. One can check that the solution to \zerofluxBB\ is given by
\eqn\bossimp{
f_{1,2,3,4,6,7} = 0.
}
We are left with one functional degree of freedom given by $f_5$. Indeed, the special thing about ${\rm T}_5$ is that it is identically zero when $\eps_{3 3} = 0$. To further constrain the energy correlator we relax the condition that $\eps_{3 3} = 0$ and again by keeping $\xi$ fixed impose the positivity condition
\eqn\bospos{
\la  {\cal E}(n) {\cal E}(\hat n) \ra_{T_{i j} \eps^{i j}(q^0)} \geq 0.
}
We first imagine that the energy correlator is finite for all values $0 < \xi \leq 1$. For example, we can choose the following polarization tensor $\eps_{11} = - \eps_{13} = 1$, $\eps_{12} = \eps_{21} = \alpha$ where $\alpha$ is real. It produces the following form that should be sign-definite for any $n_1$ and $n_2$
\eqn\quadrform{
(n^3)^2 - (n^1)^2 - 2 \alpha n^1 n^2.
}
The only way to make this form sign-definite for arbitrary $\alpha$,$n^1$ and $n^2$ such that $(n^1)^2 + (n^2)^2 \leq 1 - (n^3)^2$ is to set $n^3 = \pm 1$. One can check that the same is true for the most general polarization tensor. Since we are not considering the case when detectors are on top of each other we are left with the only solution, $n^3 = - 1$, when detectors are triggering particles propagating in the opposite directions.
In the covariant language it corresponds to $\xi = 1$.

At this moment it is clear that if we assume the finiteness we have to conclude that the two-point energy correlator is identically zero. Indeed, we cannot have a smooth function that is zero everywhere but a single point. As we explain below it is not consistent with the momentum conservation and the fact that the one-point energy correlator is nonzero.

In any CFT a flow of energy is accompanied by a flow of momentum as if it was carried by massless particles \HofmanAR\
\eqn\massless{
\vec P(n) = \vec n {\cal E}(n).
}
Consider again the reference frame where $\vec q =0$. If one detector triggers an energy flow and, therefore, a momentum flow, by momentum conservation there should be a nonzero flow of energy in other directions as well so that the two-point energy correlator cannot be identically zero at non-coincident points. Thus, we have a contradiction and we are compelled to conclude that in CFTs with finite energy correlators the three-point function of the stress tensor cannot be purely bosonic.

Let us now relax the finiteness condition and allow for integrable singularities of the type described above. Our analysis of positivity above is valid only for the points where the energy correlator is finite. We concluded that at those points the energy correlator is necessarily zero. Thus, the energy correlator is possibly nonzero only in the set of isolated points where it is singular.\foot{We freely use the term singular for the cases when the energy correlator is not regular, for instance, when it is equal to $\delta(0)$.} Combined with our assumption that energy correlators are distributions it implies that at each of those points the functions of the cross ratio $f_j(\xi)$ are given by $\sum_{n=0}^{N} c^n_j \delta^{(n)}(\xi - \xi_i^*)$ with some finite $N$ (see e.g. \vladimirov).\foot{ $\delta^{(n)}(x)$ stands for the $n$-th derivative of the  delta function.} In appendix C we show that only the $\delta(1- \xi)$ term is consistent with the positivity of energy correlators. Thus, in the $\vec q = 0$ reference frame the two-point energy correlator takes the form
\eqn\energycorrtwop{
\la {\cal E}(n_1) {\cal E}(n_2)\ra_{T_{i j}\eps^{i j} (q^0)} = (q^0)^2 \left( a + b {\eps_{i j} \eps^*_{i k} n_1^j n_1^k \over \eps_{i j} \eps^*_{i j}}+ c {\eps_{i j} \eps^*_{k l} n_1^i n_1^j n_1^k n_1^l \over \eps_{i j} \eps^*_{i j}} \right)  \delta(1 + \vec n_1 . \vec n_2).
}

Due to momentum conservation both detectors trigger the same momentum at each event and, thus, the same energy \massless.  We conclude that the following should be true
\eqn\headcoll{
\int_{\delta \Omega} d \Omega_{n_2} \la {\cal E}(n_1) {\cal E}(n_2) \ra_{T_{i j} \eps^{i j}(q^0)} = {q^0 \over 2}  \la {\cal E}(n_1) \ra_{T_{i j} \eps^{i j}(q^0)}
}
where ${q^0 \over 2}$ is the energy measured by each detector in every event and we integrate over a small region $\delta \Omega$ around the point $\vec n_1 \vec n_2 = -1$. Together with the previous constraints it fixes the two-point energy correlator to be the one of the free boson theory\foot{At this point we used $\int_0 d x \delta(x) = {1 \over 2}$.}
\eqn\totalansw{
 \la {\cal E}( n_1) {\cal E}( n_2) \ra_{T_{i j} \eps^{i j}(q^0)} = {q^0 \delta (1+ \vec n_1 . \vec n_2)\over 2 \pi}  \la {\cal E}( n_1) \ra_{T_{i j} \eps^{i j}(q^0)} .
}
In the covariant language $\xi = 1$ corresponds to $n_2^{\mu} \propto  q^{\mu} - {q^2 \over 2 q.n_1} n_1^{\mu}$.

We can proceed and fix higher point energy correlators using the total flux bound. Notice that for any $\vec n_1 \neq \vec n_2 \neq \vec n_3$ we always have a pair of detectors for which $\vec n_i . \vec n_j \neq -1$. Without a loss of generality we can think of them as being $\vec n_2$ and $\vec n_3$. But then we have
\eqn\flowconstr{\eqalign{
0 &\leq \int_{\delta \Omega} d \Omega_{\vec n_1} \la {\cal E}(n_1) {\cal E}(n_2)  {\cal E}(n_3)\ra_{T_{i j} \eps^{i j}(q^0)} \leq q^0 \la {\cal E}(n_2)  {\cal E}(n_3)\ra_{T_{i j} \eps^{i j}(q^0)} = 0
}}
for any $\delta \Omega$ and the only solution for this condition is that the three-point energy correlator is identically zero.\foot{To eliminate exotic terms like derivatives of the delta function one can consider \flowconstr~ with a nontrivial measurement function.} Analogously, we can proceed to higher point functions and conclude that all of them are zero at non-coincident points
\eqn\zerohigher{
\la {\cal E}(n_1) {\cal E}(n_2) ... {\cal E}(n_k)\ra_{T.\eps.\eps(q)} = 0, ~~ k>2.
}
This is exactly what we get in the theory of free boson. Thus, we conclude that all energy correlators in a CFT with the three-point function being purely bosonic are identical to the ones in the theory of free boson.

In the case of free theories all energy correlators with more than two detectors are zero because there is no particle production in the theory. In an abstract CFT the lack of particle production could be defined in this way. The two-point energy correlator is given by the delta function in free theories because the state created by the stress tensor is a state of two freely propagating particles. In an abstract CFT the fact that the two-point energy correlator is given by the delta function is, hence, the analog of the S-matrix triviality.

\subsec{Purely Vector Structure and ${a \over c} = {31 \over 18}$}

Now we switch to the maximum value of ${a \over c}$ and assume that the three-point function of the stress tensor is given by the purely vector structure $n_v \neq 0$, $n_b = 0$, $n_f = 0$. The one-point energy correlator takes the form
\eqn\onepointvector{\eqalign{
&\la {\cal E}(\hat n) \ra_{T.\eps.\eps(q)} ={5 \over 16 \pi} {(q^2)^4 \over (q.\hat n)^3} \left(2 - 4 {q^2 \eps.\hat n \eps^*.\hat n \over  q.\hat n^2 \eps . \eps^* } + {q^4 \eps.\hat n^2 \eps^*.\hat n^2 \over q.\hat n^4 (\eps . \eps^*)^2 } \right).
}}
And we repeat exactly the same type of analysis as described above for the case of boson. Namely we choose  $q = (q^0 , \vec 0)$ and $\hat n = (1,0,0,1)$. The condition that the one-point function \onepointboson\ is zero becomes $\eps_{1 2}=0$, $\eps_{11} = \eps_{22}$.

The zero flux constraint  \zeroen\  allows us to fix five out of seven functions
\eqn\genvec{\eqalign{
f_{7} &= 0, \cr
\xi f_1 &= - 4 \xi (1 - 2 \xi)^2 (\xi^2 - \xi -1) f_4 + (4 \xi^3 -6 \xi^2 +1) f_6, \cr
f_2 &= 24 \xi^2 (1-\xi) (2 \xi - 1)  f_4 + 2(6 \xi^2 - 6 \xi + 1)f_6, \cr
\xi f_3 &= -4 \xi (2 \xi - 1)^2 f_4 + 2 (2 \xi - 1) f_6, \cr
f_5 &= 20 \xi^3(1-\xi)  f_4 + 3 \xi (2 \xi -1) f_6.
}}
We are left with two functional degrees of freedom given by $f_4$ and $f_6$. In addition to this there is another solution localized at $\xi = 1$.
We can relax $\eps_{1 2}=0$, $\eps_{11} = \eps_{22}$ and impose the positivity of the energy correlator. The solution exists only when detectors are at the opposite points of the celestial sphere and the energy correlators are distributions and not functions.

The final result is that the energy correlators take the form
\eqn\finalvec{\eqalign{
\la {\cal E}( n_1) {\cal E}( n_2) \ra_{T_{i j} \eps^{i j}(q^0)} &= {q^0 \delta (1+ \vec n_1 . \vec n_2)\over 2 \pi}  \la {\cal E}( n_1) \ra_{T_{i j} \eps^{i j}(q^0)},\cr
\la {\cal E}(n_1) {\cal E}(n_2) ... {\cal E}(n_k)\ra_{T.\eps.\eps (q)} &= 0, ~~ k>2.
}}
Again the CFT with the extremal value of ${a \over c}$ happens to have trivial scattering observables.

\subsec{Purely Fermionic Structure and ${a \over c} = {11 \over 18}$}

It is curious that for the intermediate ${a \over c}$ the argument is not powerful enough to fix energy correlators completely.
To use the zero flux constraint we have to set $n_b n_f n_v = 0$. Here we consider the case when $n_f \neq 0$ and $n_b = n_v = 0$ which corresponds to ${a \over c} = {11 \over 18}$. The one-point function takes the form
\eqn\onepointfermion{\eqalign{
&\la {\cal E}(\hat n) \ra_{T.\eps.\eps(q)} ={5 \over 4 \pi} {(q^2)^4 \over (q.\hat n)^3} \left({q^2 \eps.\hat n \eps^*.\hat n \over  q.\hat n^2 \eps . \eps^* } - {q^4 \eps.\hat n^2 \eps^*.\hat n^2 \over q.\hat n^4 (\eps . \eps^*)^2 } \right).
}}
In the $\vec q = 0$ reference frame it is zero if $\eps_{13} = \eps_{23} = 0$. The zero flux constraint
eliminates five out of seven functions
\eqn\zerofer{\eqalign{
f_{4} &= f_{7} = 0, \cr
f_1 &= - 2 \xi (1-\xi) f_3, \cr
f_2 &= - f_6 +3 \xi (2 \xi - 1)  f_3, \cr
f_5 &= 2\xi ([2 \xi -1 ] f_6 + \xi  f_3) .
}}
There are two additional solutions localized at $\xi = {1 \over 2}$ and $\xi = 1$.

The positivity constraint again does not have solutions in the space of functions. However, if we admit distributions then we found
two solutions to the positivity condition: $\xi^* = {1 \over 2}$ and $\xi^* = 1$.
\ifig\fermions{In the case of purely fermionic structure the zero flux constraint, positivity and the total flux bound are consistent with nonzero energy flux at blue points {\it in every event and for arbitrary polarization tensor}. For each pair of blue points $\vec n_i . \vec n_j = 0,-1$. For any seven points on $S^2$ there is a pair of points such that the angle between them is neither ${\pi \over 2}$, nor $\pi$.} {\epsfxsize1.4in\epsfbox{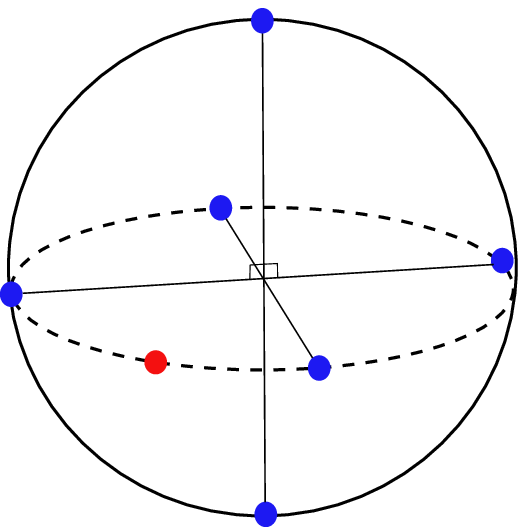}}
Using the total flux bound we see that energy correlators with more than $6$ detectors have to be identically zero (see \fermions).  Indeed, starting from $7$ detectors on the sphere there always exists a pair of points such that $\vec n_i \vec n_ j \neq 0, -1$. Nevertheless, solely from the general arguments we are not able to restrict energy correlators to the free fermion value.

\newsec{Conclusions and Summary of Results}

In all known examples of interacting CFTs all possible structures \twoandthree~are generated in
the three-point function of the stress-energy tensor. A simple way to understand this is to think about the one-point
energy correlator. The presence of all structures is translated to the statement that energy flux on the
sphere at infinity is never zero. The intuitive picture behind this is that in an interacting theory we have production
of excitations that propagate in all possible directions. Thus, the fact that energy flux is somewhere zero suggests that there
is no particle production in the theory or that it is free.

In this note we analyzed this idea using energy correlators in parity invariant four-dimensional CFTs. We showed that the {\it positivity} and {\it finiteness} of energy correlators imply that there are no CFTs with the three-point function of the stress tensor \twoandthree, which is purely bosonic, purely fermionic or purely vector.

Relaxing the finiteness as described in section $2.6$  we showed that when ${a \over c} = {1 \over 3}$ (purely bosonic structure) and ${a \over c} = {31 \over 18}$ (purely vector structure) energy correlators are those of free theories. In the case of purely fermionic structure energy correlators are very constrained as well but the positivity constraint by itself is not enough to fix them to those of the free fermion theory.

If the qualitative picture drawn above is correct it should be possible to prove a stronger statement: any CFT with $n_b n_f n_v = 0$ is necessarily free. We are not aware of any counterexample to this claim. In this form the idea can be tested in odd dimensions as well where there is no notion of $a$ and $c$ but the notion of bosonic, fermionic and vector structures are well-defined \ZhiboedovBM.

The situation here is reminiscent of the one with local operators. When the scaling dimension of an operator with spin saturates the unitarity bound the operator becomes a conserved current. Here we observe that when ${a \over c}$ saturates the unitarity bound something special happens as well. In our case the form of energy correlators suggests that higher spin symmetry emerges for extremal ${a \over c}$ \GiombiMS. 

Another example of a similar phenomenon appeared recently in the studies of crossing symmetry \RattazziPE~where interesting bounds in the space of conformal dimensions were observed \ElShowkHT, \BeemQXA. Moreover, theories that saturate these bounds were found to be very special. In \ElShowkHT~it was observed that the 3d Ising model is located on the boundary of the allowed parameter space of conformal dimensions. In \BeemQXA~4d ${\cal N}=4$ SCFTs were studied and it was conjectured that the extremal spectrum is realized by ${\cal N}=4$ SYM at the S-duality invariant value of the complexified gauge coupling. In the case of ${\cal N}=4$ SYM the theory is known to be integrable. It is not known if the same is true about the 3d Ising model \refs{\Istrail,\dotspol,\DistlerRR}.

It would be interesting to extend the present analysis in the following ways:

  \item{-} include parity-violating structures,
  \item{-} consider arbitrary spacetime dimensionalities,
  \item{-} analyze the constraints from the positivity of the two-point energy correlators without any additional assumptions,
% \item{-} analyze how these constraints arise in the bulk;
  \item{-} understand if it is possible to have nontrivial correlation functions but trivial energy correlators in CFTs.

From our analysis it is clear that in higher dimensions the positivity constraint works in a similar fashion.
Indeed, the functional freedom is bounded to eight functions of one cross ratio, while the parameter space of possible polarization tensors and detector positions for which the cross ratio $\xi$ is fixed is growing. On the other hand, it is clear that there are tensor structures that are manifestly positive-definite, thus, positivity by itself is not powerful enough to rule out nontrivial CFTs in higher dimensions.

Our analysis could be useful to extend the argument of \MaldacenaJN\ to higher dimensions. At the level of energy correlators the problem is reduced to showing that higher spin symmetry entails the stress tensor three-point function to be purely bosonic, fermionic or vector.

Imagine for a moment that trivial energy correlators imply trivial correlation functions. It would mean something very unusual for higher dimensional theories: the three-point function of the stress tensor fixes the theory completely. In AdS it means that the three-point amplitude of gravitons fix the whole S-matrix! It would be interesting to understand if analogous statements exist in flat space. The more general question along these lines would be what are the allowed three-point amplitudes and what do they say about the underlying theory? At the level of the on-shell amplitude bootstrap there seems to be no problem; however, it could be that there are additional subtle causality and unitarity constraints \AdamsSV.

We hope that further analysis of energy correlators can lead to additional interesting insights about the higher-dimensional
conformal field theories as well as their gravitational duals.

{\it Acknowledgments:} We are thankful to A. Belitsky, S. Caron-Huot, A. Dymarsky, S. Hohennegger, E. Sokatchev for interesting discussions on this and related topics. We are especially grateful to G. Korchemsky and J. Maldacena for insightful discussions and comments on the manuscript. The work of AZ was supported in part by the US National Science Foundation under Grant No. PHY-0756966.

\appendix{A}{Different Choices of Detectors}

Here we explain equivalence of the detector definitions \detA\ and \detB\ using conformal transformation considered in \refs{\CornalbaFS,\HofmanAR}.
The transformation goes as follows
\eqn\transform{\eqalign{
y^+ &= - {1 \over x^+},~y^- = x^- - {x_1^2 + x_2^2 \over x^+}, \cr
y_1 &= {x^1 \over x^+},~y_2 = {x^2 \over x^+},
}}
where $y^{\pm} = y^0 \pm y^3$ and $x^{\pm} = x^0 \pm x^3$. Both large $r$ and $(x. \bar n)$ limits correspond to $x^+ \to \infty$ so that after the conformal transformation detectors are inserted at the null plane $y^+ = 0$. If we consider the contraction $T_{\mu \nu} \zeta^{\mu \nu}$ then after the transformation the polarization tensor takes the form
\eqn\changeof{
\hat \zeta^{\hat \mu \hat \nu} = \zeta^{\mu \nu} {\pa y^{\hat \mu}  \over \pa x^{\mu}} {\pa y^{\hat \nu} \over x^{\nu}}.
}
The crucial point is that $ {\pa y^{+,1,2}  \over \pa x^{\mu}}|_{y^+ = 0} = 0$, so that after taking the limit arbitrary choice of the polarization tensor $\zeta^{\mu \nu}$ is related to the expectation value of $T_{--}(y^+=0,y^-,y^1,y^2)$. One can use this fact to argue that the energy and momentum flows in CFTs are related to each other as in theories with massless particles. The nonzero components take the form
\eqn\jacobian{
{\pa y^- \over \pa x^0} = {2 \over 1 + n_3},~~{\pa y^- \over x^i} = -{2 n^i \over 1 + n_3}.
}
Thus, different choices of polarization tensors for detectors produce
\eqn\detector{
\lim_{r \to \infty} r^2 T_{\mu \nu} (x) \zeta^{\mu \nu} \to {1 \over (1 + n^3)^2} {4  \zeta.n.n\over (1 + n^3)^2} T_{--}(y^+=0,y^-,y^1,y^2),
}
where we separated $\lim_{r \to \infty} r^2 (y^+)^2 = {1 \over (1 + n^3)^2}$ coming from the Jacobian $(y^+)^2$ which appears in addition to the change of polarization when we apply a conformal transformation to the primary field.

The detector choice of Hofman and Maldacena $T_{0 i} n^i$ produces
\eqn\hmchoice{
\zeta^{{\rm HM}}.n.n = 1,
}
if we consider instead $T_{\mu \nu} \bar n^{\mu} \bar n^{\nu}$ we get the factor $(n. \bar n)^2 = 4$ which explains an extra factor of ${1 \over 4}$ in \detB\ relative to \detA. Another ${1\over 4}$ factor comes from $\lim_{(x. \bar n) \to \infty} (x.\bar n)^2 (y^+)^2 = 4 {1 \over (1 + n^3)^2}$.

\appendix{B}{Tensor Structures for The Two-point Energy Correlator}

The tensor structures that appear in the two-point energy correlator take the following form

\eqn\structures{\eqalign{
{\rm T}_1&=1,~{\rm T}_2={\eps.n_2 \eps^* .n_1 + \eps.n_1 \eps^* .n_2 \over n_1. n_2 \eps.\eps^*}, \cr
{\rm T}_3&={q^2 \over \eps.\eps^*} \left( {\eps.n_2 \eps^* .n_2 \over q.n_2^2} + {\eps.n_1 \eps^* .n_1 \over q.n_1^2} \right), \cr
{\rm T}_8&={|\eps.n_2 \eps.n_1|^2 \over (\eps.\eps^*)^2 (n_1 .n_2)^2},~{\rm T}_4={q^4 \over  (\eps.\eps^*)^2 } \left( {|\eps.n_2|^4 \over q.n_2^4}+ {|\eps.n_1|^4 \over q.n_1^4}\right), \cr
{\rm T}_5&={(\eps.n_2 \eps^* .n_1)^2 + (\eps.n_1 \eps^* .n_2)^2 \over (\eps.\eps^*)^2 (n_1 .n_2)^2}, \cr
{\rm T}_6&=q^2 {\eps.n_2 \eps^* .n_1 + \eps.n_1 \eps^* .n_2 \over  (\eps.\eps^*)^2 n_1.n_2}\left( {|\eps.n_2 |^2 \over q.n_2^2}+ {|\eps.n_1 |^2 \over q.n_1^2} \right), \cr
{\rm T}_7&= i q^2 {\eps.n_2 \eps^* .n_1 - \eps.n_1 \eps^* .n_2 \over  (\eps.\eps^*)^2 n_1.n_2} \left( {|\eps.n_2|^2 \over q.n_2^2}- {|\eps.n_1|^2 \over q.n_1^2} \right).
}}
Curiously, not all of them are linearly independent in $d=4$. We found that
\eqn\relation{\eqalign{
{1 - \xi \over 2 \xi}{\rm T}_{1} + {1 - 2 \xi \over 2 \xi}{\rm T}_{2} - { 1 \over 4 \xi^2}{\rm T}_{3} - {1 \over 2} {\rm T}_{5} +{\rm T}_{8}=0.
}}
This relation is special to four dimensions and can be equivalently written as ${\rm det} (\alpha_i . \alpha_j) = 0$ where $\alpha = \{ q, n_1 ,n_2, \eps, \eps^* \}$. We use \relation\ to eliminate ${\rm T}_{8}$.

It is also trivial to go from the polarization tensor $\eps^{\mu} \eps^{\nu}$ to $\eps^{\mu \nu}$ used sometimes in the bulk of the paper. Indeed, each structure above could be rewritten in terms of $\eps^{\mu \nu}$ and $(\eps^*)^{\mu \nu}$ in a unique way. The opposite is also true.

Imagine the above is not correct and there are two different $\eps^{\mu \nu}$ structures that produce the same $\eps^{\mu} \eps^{\nu}$ structure. It means that there exists a tensor $M_{\mu \nu \rho \sigma}$ such that
\eqn\equivalence{\eqalign{
M_{\mu \nu \rho \sigma} \eps^{\mu \nu} (\eps^*)^{\rho \sigma} &\neq 0, \cr
M_{\mu \nu \rho \sigma} \eps^{\mu} \eps^{ \nu} (\eps^*)^{\rho} (\eps^*)^{ \sigma} &= 0.
}}
The building blocks for $M_{\mu \nu \rho \sigma}$ are $\delta^{\mu \nu}$, $n_1^\mu$, $n_2^\mu$. It is clear that such a tensor does not exist.

\appendix{C}{Constraints on the Possible Delta-function Terms}

In the bulk of the paper we concluded that the two-point energy correlator is nonzero at isolated points $\{ \xi^*_i \}$ where it is given
by a finite sum of the delta function and its derivatives $\sum_{n=0}^{N} c_n \delta^{(n)}(\xi - \xi^*)$. Here we demonstrate that only the $\delta(1-\xi)$ term is allowed by the integrated version of the positivity constraint.

We consider the integrated version of positivity over a small region around $\xi_i^*$
\eqn\twopointconstr{
 \lim_{\delta \xi \to 0}\int_{\xi_j^* -\delta \xi}^{\xi_j^* + \delta \xi} d \xi  ~g(\xi) {q^2 \over (n_1 . n_2)^3} \sum_{i=1}^{7} {\rm T}_{i} f_{i} (\xi) \geq 0,
}
where $g$ stands for some {\it non-negative} measurement function.
We choose measurement functions to be $g_n(\xi) = \left(\delta \xi - (\xi - \xi^*) \right)^n$. First, we consider $g_N$. Notice that in the small $\delta \xi$ limit only the term $\delta^{(N)}(\xi - \xi_i^*)$ contributes with all derivatives acting on $g_N$. Hence, the positivity condition takes the form
\eqn\twopointconstrB{
(-1)^N {q^2 \over (n_1 . n_2)^3} \sum_{i=1}^{7} {\rm T}_{i} c^{N}_{i} \geq 0,
}
and analysis is identical to the one we had before when we assumed energy correlators to be regular. In this manner for all $\xi^* \neq 1$ we set $c^N_{i} = 0$. We can repeat the procedure for $g_{N-1}$ until we eliminate all $n \geq 0$ and conclude that the energy correlator is zero away from $\xi = 1$.

We end up with the detector two-point function having the following form $\sum_{n=0}^{N} c_n \delta^{(n)}(1 - \xi)$. In the $\vec q =0$ reference frame it becomes
\eqn\twopointgen{\eqalign{
\la {\cal E}( n_1) {\cal E}( n_2)\ra_{T_{i j}\eps^{i j} (q^0)} = (q^0)^2 \sum_{i=1}^{7} \sum_{n=0}^{N} {\rm T}_i  c^n_i \delta^{(n)}(1 + \vec n_1 . \vec n_2)
}}
Let us check positivity of \twopointgen~ by integrating it $\lim_{\eps \to 0}\int_{-1}^{-1+\eps} d \vec n_1. \vec n_2 $ with non-negative measurement functions chosen as follows $g_n =\left( \eps \pm (1+ \vec n_1 . \vec n_2) \right) \left( \eps + (1+ \vec n_1 . \vec n_2) \right)^{n-1}$. We start with $g_N$ and the positivity problem is identical to the one for the one-point function since the only term that contributes comes from $\delta^{(N)}$ term that acts solely on the measurement function. By choosing $+$ or $-$ in the measurement function we effectively change the sign of constants $c_i^N$. It is easy to show that the only solution that is consistent with positivity for both signs is $c^N_i = 0$. We then repeat the same procedure for all $g_n$ with $n \geq 1$. It is clear that $n=0$ is special and we do not have the plus-minus choice in the measurement function. In this way the two-point energy correlator is proportional to $\delta (1 + \vec n_1 . \vec n_2)$.

\listrefs

\bye